\documentclass[aps,prl,twocolumn,showpacs,superscriptaddress,groupedaddress]{revtex4}
\usepackage{graphicx}
\usepackage{dcolumn}
\usepackage{bm}
\usepackage{amssymb}
\usepackage{color}

\usepackage[latin9]{inputenc}
\usepackage{esint}
\usepackage{bbm}
\usepackage{amsfonts}
\usepackage{mathrsfs}
\usepackage{ulem}

\begin{document}

\title{Geometric Measure of non-Commuting Simultaneous Measurement based on $K$-Means  Clustering}

\author{Yang Yang}
\author{Wei Cui}
\email{aucuiwei@scut.edu.cn}
\address{School of Automation Science and Engineering, South China University of Technology, Guangzhou 510641, China}

\date{\today}

\begin{abstract}
Considering the  simultaneous   measurement of non-commuting observables, we define a geometric measure for the degree of non-commuting behavior of quantum measurements coming from the initial and final states of the measurements. The rationality of our geometric measure is demonstrated and the application of it is presented. The connection between our measure and Heisenberg's uncertainty principle is discussed as well. Our work deepens the understanding of quantum non-commuting measurement.
\end{abstract}


\maketitle


Research on quantum measurement theory has a long history and a rough architecture of it has already been built \cite{Braginsky1992,Wiseman2009,Jacobs2014,Arthurs1965,Busch1985,Stenholm1992,Jordan2005}. Promoted by the advancement of quantum experimental technology \cite{Blais2004,Campagne-Ibarcq2014,Xiang2013,Slichter2012,Hatridge2013,Murch2013}, some new sub-areas of quantum measurement theory have emerged and earned many attentions recently; for instance, quantum non-commuting measurement \cite{Wei2008,Ruskov2010,Ruskov2012,Hacohen-Gourgy2016,Atalaya2018}. Heisenberg's uncertainty principle points out that non-commuting observables such as position and momentum can not be precisely measured simultaneously. What's more, such measurements is impossible with projective measurement, while it has been proved that simultaneous non-commuting measurements can be performed using continuous weak quantum measurements \cite{Arthurs1965}. Moreover, continuous quantum measurement will eventually convert to projective measurement when the measurement strength increases to a certain extent, thus when a simultaneous measurement of non-commuting observables is given, whether it can be physically realized is a problem worth considering.

The theoretical analysis of the simultaneous measurement of non-commuting observables can be traced back to the middle and late 20th century \cite{Arthurs1965,Busch1985,Stenholm1992,Jordan2005}. In the last decade, it re-attracted many attentions thanks to the development of the quantum experimental technology \cite{Wei2008,Ruskov2010,Ruskov2012,Hacohen-Gourgy2016,Atalaya2018,Garcia-Pintos2016,Garcia-Pintos2017}. Ref.~\cite{Wei2008} analyzed the statistics of the measured outputs and the fidelity of system state monitoring via the measured outputs when non-commuting measurement is performed on a qubit. The dynamics caused by the simultaneous measurement of non-commuting observables has been theoretical discussed in Ref.~\cite{Ruskov2010}, and experimentally demonstrated in Ref.~\cite{Hacohen-Gourgy2016}.
In Ref.~\cite{Atalaya2018}, the temporal correlation of the two output signals of quantum non-commuting measurement has been discussed and further applied to quantum parameter estimation.

Up to now, the measure for the degree of non-commuting behavior of quantum measurements has not been discussed in relevant papers. Therefore this paper considers defining such a measure, and further determining whether a quantum measurement can be  physically realized based on this measure. With the measure for non-commutability of quantum measurement, many related work in the field of quantum non-commuting measurement may be able to make further progress: when analyzing data coming from experiments or simulations, this measure can help to understand the phenomenon presented by the data, and can also provide a new idea for data processing; this measure is also expected to provide guidance for the design of the experiments. Most importantly, this measure is looked forward to help to {deep} the understanding of quantum non-commuting measurement, thereby driving the discovery of new physical phenomenon.



Different from the classical world, quantum measurements cause dynamical changes\cite{Braginsky1992,Wiseman2009,Jacobs2014}. To construct a general measure, the measure is expected not to refer to any specific mathematical representation of the dynamics, therefore the data that can be used as input is limited to the initial and final states of the measurement. Moreover, the dynamics induced by a specific quantum measurement is uncertain, which means that quantum states after measurement need to be represented using a set of density matrices. In this way, when measuring the non-commutability of a quantum measurement, $N$ (sufficiently large) copies with a same initial state $\rho_0$ are prepared, and the measurement is performed for each copy, the set of the obtained $N$ final states is denoted as $R_f$. Thus, the data that can be used as inputs are $\rho_0$ and $R_f$.

The dynamics of quantum continuous weak measurements considered in this paper can be described by the following stochastic master equation \cite{Wiseman2009}:
\begin{eqnarray}
d\rho&=&-\kappa_0\big[\sigma_\phi,[\sigma_\phi,\rho]\big]dt \nonumber\\&&+\sqrt{2\kappa_0}\big[\sigma_\phi\rho, \rho\sigma_\phi-2{\rm Tr}(\sigma_\phi\rho)\rho\big]dW,
\label{eq1}
\end{eqnarray}
where $dW$ is the standard Wiener process, $k$ and $\sigma_\phi$ represent the measurement strength and the measured observable, respectively. Furthermore, according to Reference \cite{Hacohen-Gourgy2016} , the dynamics of the systems can be described using the following stochastic master equation when the non-commuting observables are simultaneously measured
\begin{eqnarray}
d\rho&=&\sum\limits_{i=1}^{2}{-\kappa_i\big[\sigma_{\phi i},[\sigma_{\phi i},\rho]\big]dt}\nonumber \\&&+ \sum\limits_{i=1}^{2}{\sqrt{2\kappa_i}\big[\sigma_{\phi i}\rho,\rho\sigma_{\phi i}-2{\rm Tr}(\sigma_{\phi i}\rho)\rho\big]dW_i}
\label{eq2}
\end{eqnarray}
where $\sigma_{\phi i}$ is the  measured observable, $k_i$ is the measurement strength, and $dW_i$ is the corresponding standard Wiener process.

Considering quantum measurements of a single observable that perform no non-commuting behavior, system states will gradually toward one of the two eigenstates of the measured observable \cite{Braginsky1992}, that is, the set of final states can be divided into two subsets corresponding to two eigenstates of the observable respectively. Furthermore, when quantum measurements of non-commuting observables are considered, there will be four eigenstates affecting the evolutions of system states. However, each eigenstate of certain observable will be closer to one of the eigenstate of the other observable when the angle $\theta$ between two observables is less than $\pi/2$, under the affect of which the set of final states can be divided into two subsets (correspond to the combinations of the eigenstate of one observable and the closer eigenstate of the other observable respectively) in addition.

Thus, a clustering method is demanded to divide the set of final states of the measurement into two subsets and the statistical characteristics of which are further analyzed to construct the measure. The $K$-means method is chosen in this paper. The $K$-means method is a prototype-based objective function clustering method that selects the sum of the Eucilidean distances of all objects to the prototype as the objective function of the optimization \cite{Lloyd1982,Jain2010}. A brief introduction to the flow and theory of the K-means method is given below.

To cluster all objects into $K$ class, first select $K$ initial particles randomly, assign each object to the particle with the smallest Euclidean distance to form $K$ clusters, and calculate the mean of each cluster as the new $K$ particles. Iterate continuously until the shutdown condition is met \cite{Lloyd1982} (the iteration is stopped when the distance between the new and old particles is less than a sufficiently small value $\lambda=0.01$ in this paper). In this way, we can easily classify all the objects into $K$ classes.

\begin{figure}
\setlength{\abovecaptionskip}{6pt} \centerline{\scalebox{0.8}[0.8]{\includegraphics{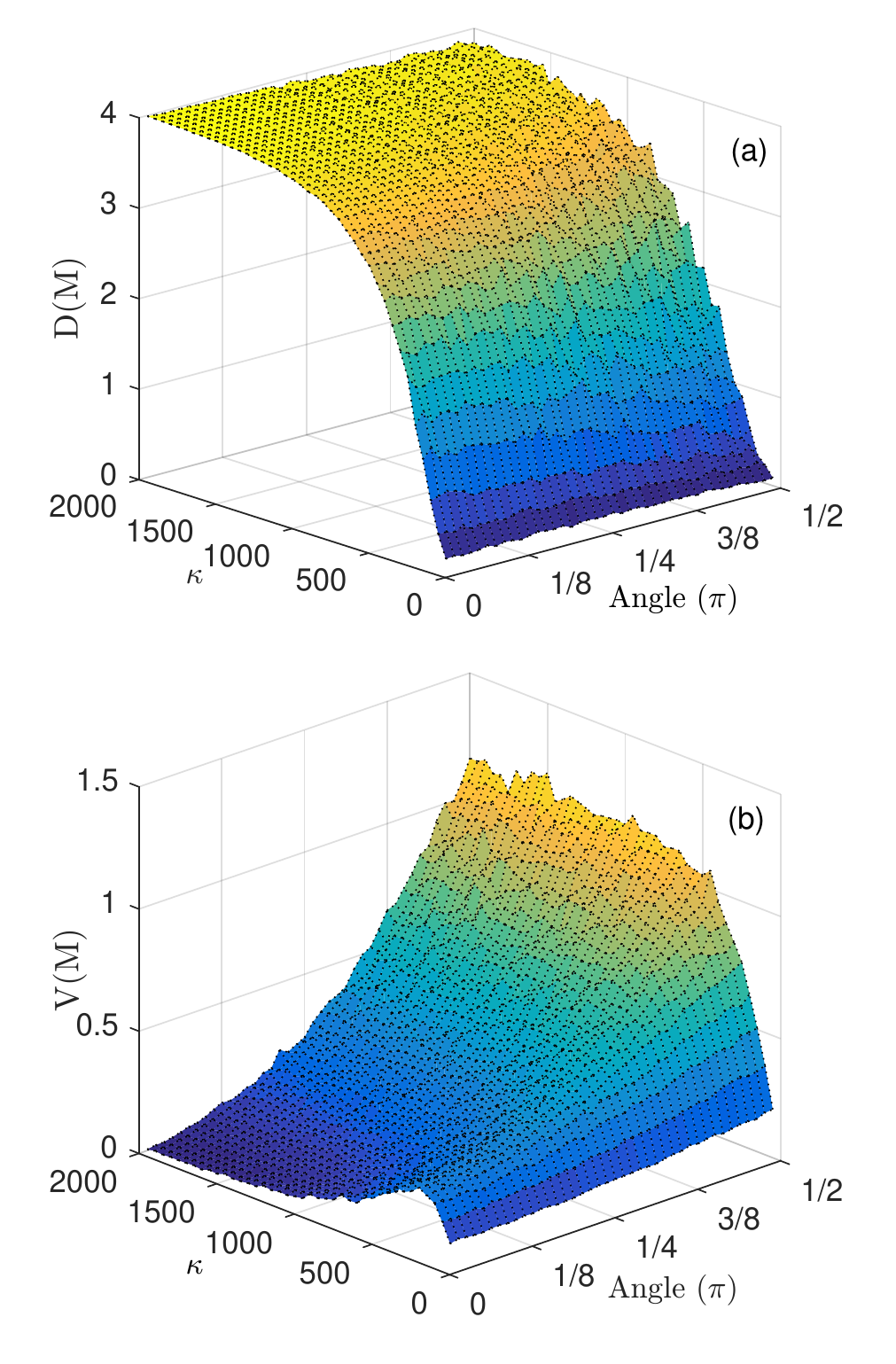}}}
\caption{(Color online) { The intermediate variable $D(M)$   and $V(M)$   as functions of measurement strength $\kappa$ and angle $\theta$ between two measurement observables ($\sigma_{\phi 1}=\sigma_z$ and $\sigma_{\phi 2}={\rm sin}\theta\sigma_x+{\rm cos}\theta\sigma_z$). The initial state $\rho_0=[0.8,0.4; 0.4,0.2]$ and the dynamics of systems are calculated by Eq.~(\ref{eq2}) with measurement duration $T=200\mu s$.}}\label{fig1}
\end{figure}

Then we will give the expression of the measure for the non-commutability of quantum measurement defined and demonstrate that it satisfies the properties to be satisfied as a measure. For $N$ (sufficiently large) copies whose initial states are all $\rho_0$, perform the measurement $M$ on them and obtain the set of final states $R_f$. $R_f$ is clustered using the $K$-means method, $K=2$, and the two subclasses $R_{f1}$ and $R_{f2}$ are obtained. The measure defined by us thus can be given below:
\begin{eqnarray}
{P}_i(M)&=&\rho_{fi}\bigg(\mathop {\arg \min}\limits_{k=1,2,...,N_i} {\sum\limits_{j=1}^{N_i}{\parallel B(\rho_{fi}(k))-B(\rho_{fi}(j))\parallel_2^2}}\bigg),\nonumber\\
D(M)&=&\sum\limits_{i=1}^{2} {{\parallel {B(\rho_0)-B({P}_i(M))}\parallel_2^2}},\nonumber\\
V(M)&=&{1\over N}\sum\limits_{i=1}^{2} {\sum\limits_{j=1}^{N_i} {\parallel B(\rho_{fi}(j))-{1\over N_i}\sum\limits_{j=1}^{N_i}{B(\rho_{fi}(j))}}\parallel_2^2},\nonumber\\
\Phi(M)&=&\alpha {V(M)\over {D(M)(4-D(M)+\gamma)}}-\beta.
\end{eqnarray}
\begin{figure}
\setlength{\abovecaptionskip}{6pt}
\centerline{\scalebox{0.6}[0.6]
{\includegraphics{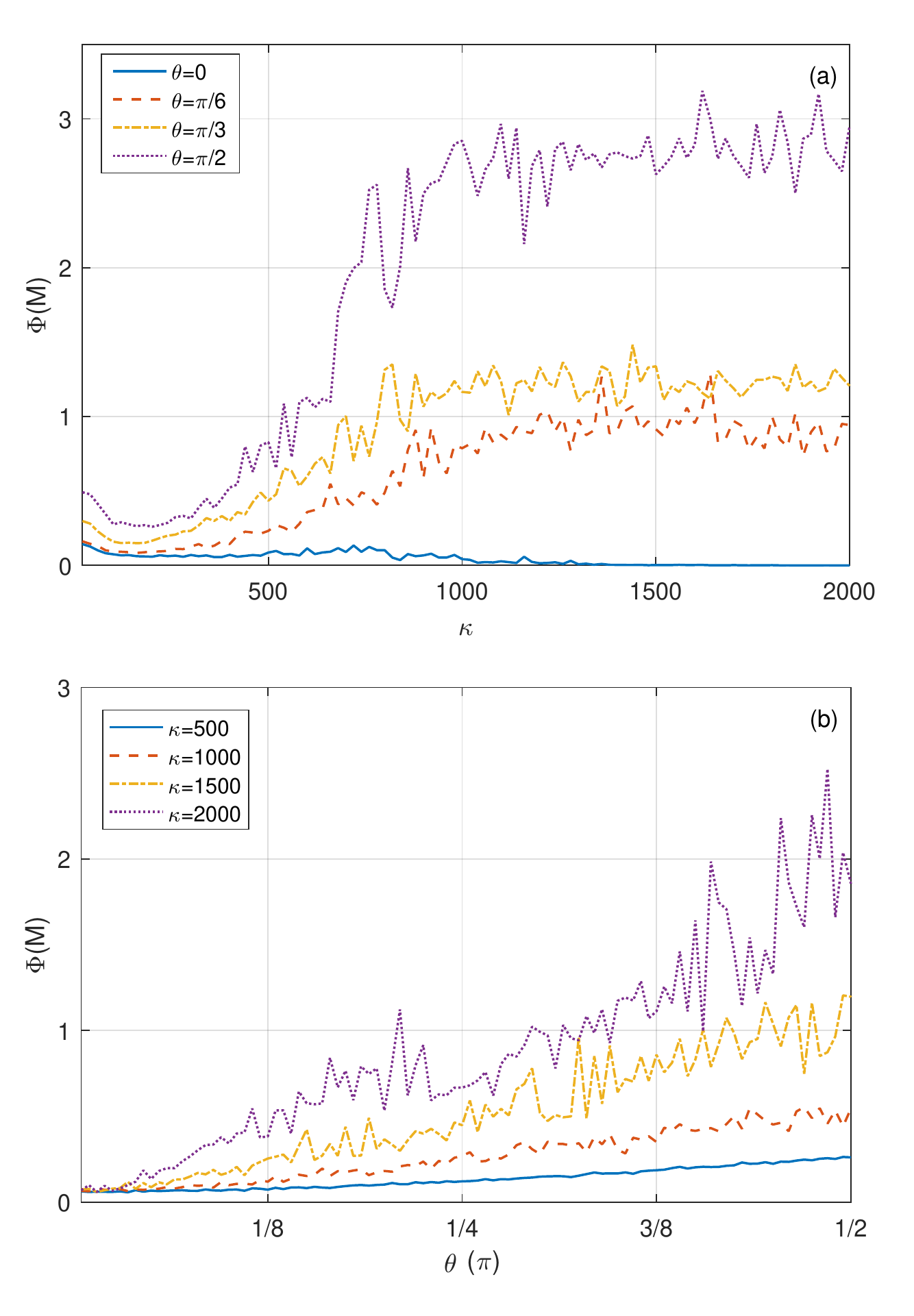}}}
\caption{(Color online)  Fig. (a) and Fig. (b) show the defined geometric measure of non-commuting simultaneous measurement $\Phi(M)$ as a function of measurement strength $\kappa$ and non-commutability angle $\theta$ for various parameters, respectively. The remaining system parameters are the same with Fig~\ref{fig1}.}\label{fig2}
\end{figure}

Here {$P_i(M),i=1,2$, is the element of certain subset that minimizes the sum of distances from other elements in the subset, and is expected to reflect the average position of subset elements, }$D(M)$ is the intermediate variable which is able to reflect the measurement strength obtained from $\rho_0$ and $P_i(M)$, {$V(M)$ is the weighted sum of variances of two subsets which is expected to reflect the angle $\theta$ between two observables, and $\Phi(M)$ is the measure constructed finally. }$N_1$ and $N_2$ are the cardinalities of $R_{f1}$ and $R_{f2}$. $\rho_{fi}(j)$ is the $j$-th element of $R_{fi}$ and $B(\rho)$ represents the Bloch vector of density matrix $\rho$. $\alpha$, $\beta$, and $\gamma$ are auxiliary parameters whose values need to be further determined under the constraints of $\alpha,\beta,\gamma>0$ and $\gamma\to0$.


Fig.~\ref{fig1} shows $D(M)$ and $V(M)$ as functions of the measurement strength $\kappa_i$ and the non-commuting angle $\theta$ between two observables, respectively. We consider the simplest case of two identical detectors: $\kappa_1=\kappa_2=\kappa$.
We can see that $D(M)$ simply increases with $\kappa$ as expected, while $V(M)$ is affected by not only $\theta$ but also $\kappa$.
 Moreover, the defined geometric measure $\Phi(M)$ as a function of the measurement strength $\kappa$  for various non-commuting angles is plotted in Fig. 2(a). Fig. 2(b) represents the evolution as a function of the non-commuting angle $\theta$ for various measurement strength $\kappa$. Other parameters in Eq.(2) are
$\alpha=1, \beta=0, \gamma=0.01$. It's obvious that the
geometric measure defined  increased with $\kappa$ as well as $\theta$ as expected, a more detailed analysis will be given below.

\begin{figure}
\setlength{\abovecaptionskip}{6pt} \centerline{\scalebox{0.68}[0.68]{\includegraphics{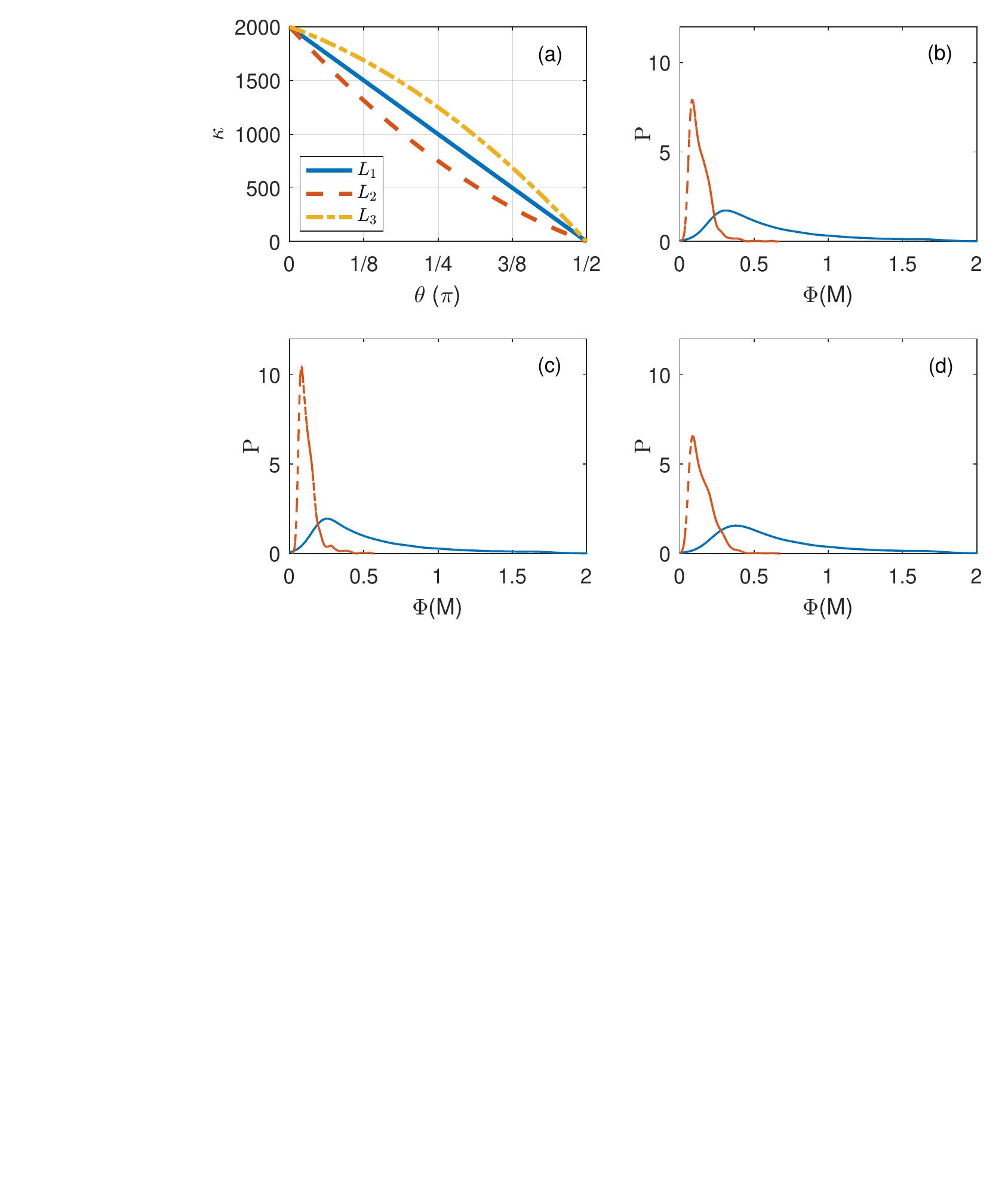}}}
\caption{(Color online) {(a): The three reference curves ($L_1$, $L_2$, and $L_3$) assumed to be closest to the real bound. Figs. (b),(c), and (d) show the two probability density curves coming from the two sets of values of $\Phi(M)$ below and above $L_1$, $L_2$, and $L_3$ , respectively. }}\label{fig4}
\end{figure}

A simple simulation experiment is done to show that our measure is a useful tool in the field of quantum non-commuting measurement in addition. Considering the dynamics described by the stochastic master equation~(\ref{eq2}), $\kappa_1=\kappa_2=\kappa$, there must exist a bound above which quantum measurement will become unable to physically realized. Specific to the case we consider, there are only two parameters that can be changed ($\kappa$,$\theta$), thus the bound can be represented by a curve in the two-dimensional coordinate system whose axes represent the two variable parameters respectively. Without loss of generality, we assume that there exist three reference curves ($L_i, i=1,2,3$) closest to the real bound and use our measure to distinguish the three reference curves and finally select the optimal one.
From our point of view, $\Phi(M)$ below and above the real bound are supposed to have most different performance. Therefore we denote the sets of the values of $\Phi(M)$ below and above certain curve $L_i(i=1,2,3)$ as $S1_{L_i}$ and $S2_{L_i}$, use the Matlab function $ksdensity$ to plot the probability density curves of $S1_{L_i}$ and $S2_{L_i}$, and simply calculate the proportion of overlapping parts of two probability density curves to determine the optimal bound curve.
Fig.~\ref{fig4} shows the three reference curves chosen and two probability density curves of them. Tab.~\ref{table1} shows the proportion of overlapping parts of the three reference curves, through which we finally select $L_1$ to be the optimal bound curve.

\renewcommand\arraystretch{2}
\begin{table}
\caption{The proportion of overlapping parts of two probability density curves}
\begin{tabular}{|p{2cm}<{\centering}|p{2cm}<{\centering}|p{2cm}<{\centering}|p{2cm}<{\centering}|}
\hline
Reference Curve & $L_1$ & $L_2$ & $L_3$ \\
\hline
Proportion & 20.08\% & 21.82\% & 21.00\%\\
\hline
\end{tabular}
\label{table1}
\end{table}
\renewcommand\arraystretch{0.5}

\begin{figure*}
\setlength{\abovecaptionskip}{6pt} \centerline{\scalebox{0.8}[0.8]{\includegraphics{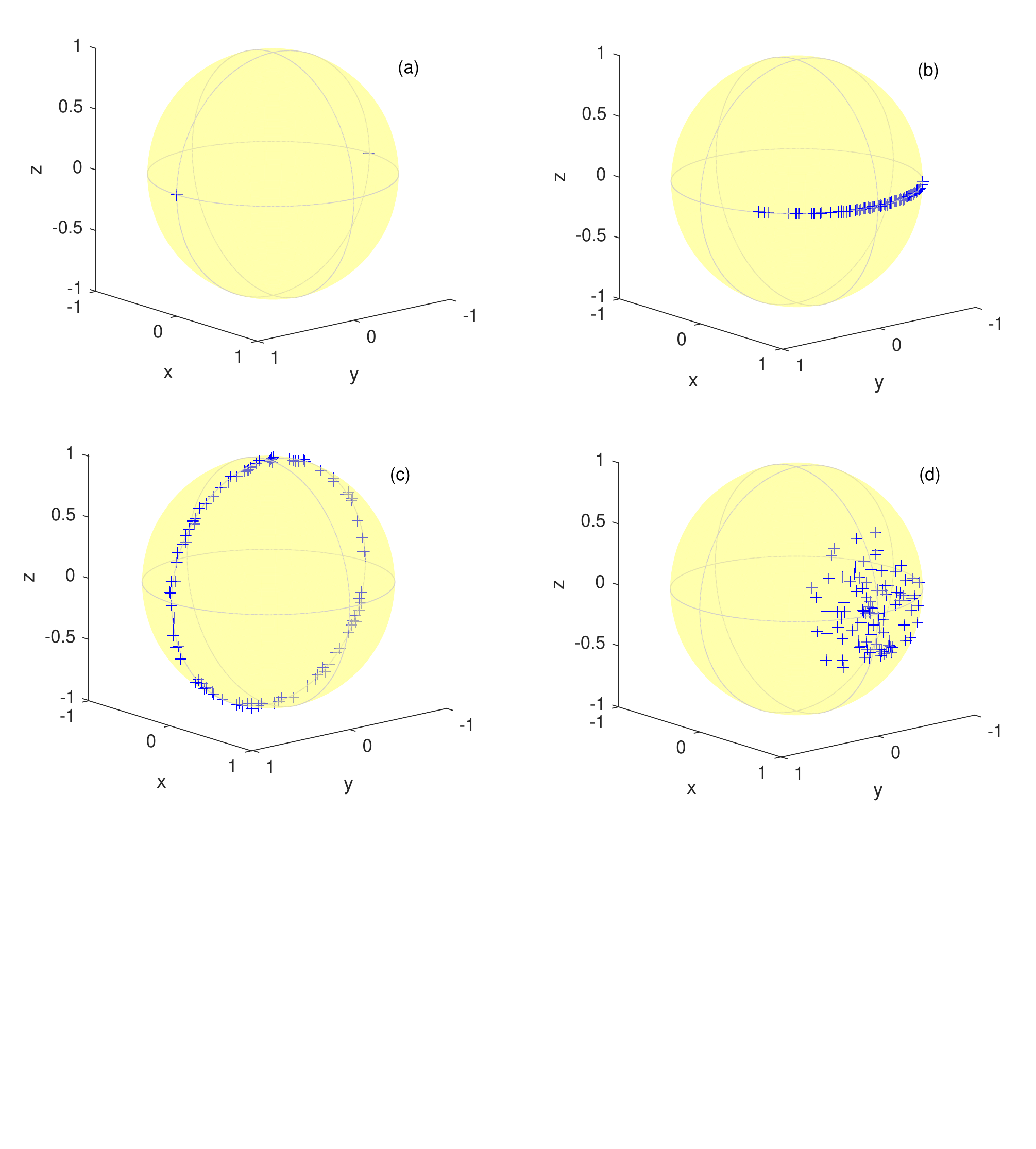}}}
\caption{(Color online) Several typical cases in quantum measurements, where (a) and (b) correspond to projection measurements of a single observable and non-commuting observables, respectively; (c) and (d) correspond to continuous weak measurements of a single observable and non-commuting observables, respectively.}\label{fig3}
\end{figure*}


Finally, let us explore whether the measure for the non-commutability of quantum measurements defined above can give reasonable results for several typical cases in quantum measurements which are presented in Fig.~\ref{fig3}, thus indicating the rationality of the measure in the aspect of physical meaning. Here, we simply use $M_{\rm proj}$, $M_{\rm weak}$, and $M_{\rm ncpr}$ to represent projection measurement, continuous weak measurement, and non-commuting projection measurement, respectively.  Firstly, the simplest projection measurement of a single observable $\sigma_z$ is considered (Fig.~\ref{fig3} (a)), after which the system will be in one of the basic states of $\sigma_z$ ($\rho_0=\left|0\right\rangle\left\langle0\right|$ and $\rho_1=\left|1\right\rangle\left\langle1\right|$), with probability $p_{0}={\rm Tr}(\rho\rho_{0})$ or $p_{1}={\rm Tr}(\rho\rho_{1})$. This means that the elements in a subset obtained by the $K$-means method can only take one of the two basic states of $\sigma_z$ (the elements of the other subset can only take the other basic state of $\sigma_z$), causing $V(M_{\rm proj})=0$ and $\Phi(M_{\rm proj})=-\beta$ gets its minimum. While there is obviously no non-commuting behavior for projection measurement of a single observable, so it's established that the measure defined works well with projection measurements of a single observable.

The influence of the increase in the angle $\theta$ between two observables on the measure defined when the measurement strength $k$ is small and fixed is considered in addition (Fig.~\ref{fig3} (b) and Fig.~\ref{fig3} (d)). In the physical sense, the non-commutability of quantum measurement in this case will definitely increase as $\theta$ increases.  For the $\Phi(M_{\rm weak})$ defined in this paper, its value is mainly determined by $V(M_{\rm weak})$ with a small $k$, and the increase of $\theta$ will make the final states of the measurement more evenly distributed on the surface formed by the initial state and the steady plane which is determined by the eigenstates of the two observables, resulting in an increase in $V(M_{\rm weak})$ and then an increase in $\Phi(M_{\rm weak})$, which is consistent with physical sense.
The last typical case considered is the projection measurements of non-commuting observables (Fig.~\ref{fig3} (c)). In this case, the measurement will become unable to physically realize if $\theta$ takes any value other than $0$, which means that the measure defined should take a value larger than a certain boundary. When the measurement strength increases to the degree of projection measurement, there is $D(M_{\rm ncpr})=4$, causing $\Phi(M_{\rm ncpr})$ to take a sufficiently large value once $V(M_{\rm ncpr})$ takes a non-zeros value. Moreover, only when $\theta=0$, $\Phi(M_{\rm ncpr})=V(M_{\rm ncpr})=0$ will be established. Through those typical cases, the rationality of the measure defined in the aspect of physical meaning is illustrated.

In conclusion, this paper propose a measure for the non-commutability of quantum measurements based on the $K$-means Clustering method and demonstrate its rationality. We further consider the application of our measure in several typical cases in quantum measurements to indicate its practicality. Our work helps to advance the understanding of quantum non-commuting measurement.
{As quantum measurement has been applied to many other fields such as quantum control and quantum state estimation \cite{Yang2018,Gong2018,Weber2014,Vijay2012,Zhang2017,Gillett2010}, the applications of non-commuting quantum measurement in these fields are of great expectations in addition \cite{Ruskov2010,Hacohen-Gourgy2016}. However, the more information is gained with non-commuting measurement, the more backaction and uncertainty exist. Thus how to choose the measurement that makes use of as much information as possible with acceptable backaction and uncertainty is of great interest and importance. Our measure is expected to help to solve this problem originally.} Moreover, we wonder whether there will exist a connection between our measure and Heisenberg's uncertainty principle. It's known that the quantum measurement will become unable to physically realize when its measure exceeds certain boundary, we are looking forward to precisely obtain this boundary through Heisenberg's uncertainty principle, which might be completed in the future works.

\begin{acknowledgements}
This work was supported by the National Natural Science Foundation of China under Grant 61873317, and by the Fundamental Research Funds for the Central Universities.
\end{acknowledgements}

\label{sec:TeXbooks}

\end{document}